\documentclass[12pt]{article}
\usepackage{amssymb}
\usepackage{mathrsfs}
\usepackage{bm}
\usepackage{graphicx}
\usepackage{subfigure}
\usepackage{hyperref}
\usepackage[utf8x]{inputenc}

\begin{document}
\begin{titlepage}
%\begin{flushright}
%v11.2
%\today
%\end{flushright}
\begin{center}
\begin{Large}
{\bf  Higher time derivatives in the microcanonical ensemble describe dynamics of flux--coupled  classical and quantum oscillators}
\end{Large}

\vskip1.0truecm
Stam Nicolis\footnote{E-Mail: stam.nicolis@idpoisson.fr; stam.nicolis@lmpt.univ-tours.fr}

{\sl CNRS--Institut Denis Poisson (UMR 7013)\\
Université de Tours, Université d'Orléans\\
Parc Grandmont, 37200 Tours, France
}

\end{center}

\vskip1truecm

\begin{abstract}

We show that it is possible to consistently describe dynamical systems, whose equations of motion are of degree higher than two, in the microcanonical ensemble, even if the higher derivatives aren't coordinate artifacts. Higher time derivatives imply that there  are more than one Hamiltonians, conserved quantities due to time translation invariance, and, if the volume in phase space, defined by their intersection, is compact, microcanonical averages can be defined and there isn't any instability, in the sense of Ostrogradsky, even though each Hamiltonian, individually,  may define a non--compact (hyper)surface. 

We provide as concrete example of these statements  the Pais--Uhlenbeck oscillator and show that it  can  describe a system that makes sense in the microcanonical ensemble. It describes two oscillators that are coupled by imposing a fixed phase difference, that thereby describes a non--local interaction between them. The consistent quantum dynamics can straightforwardly be expressed using two pairs of creation and annihilation operators, with the phase difference describing a flux, that describes the interaction. 

The properties of the action imply  that particular solutions, that would describe independent oscillators, are, in general,  not admissible.The reason is that the coordinate transformation, that would decouple the oscillators isn't a symmetry of the action--unless  a ``BPS bound'' is saturated.  Only then do they decouple. But, in these cases, the action does describe one, not two, oscillators, anyway and the higher derivative term is a coordinate artifact.

\end{abstract}
\end{titlepage}
A dynamical system, that's invariant under time translations, possesses a corresponding conserved quantity, the energy. For a system with a finite number of degrees of freedom, the equation
\begin{equation}
\label{H=E}
H(p,q)=E
\end{equation}
defines a (hyper)surface in phase space, where $p$ and $q$ denote the phase space variables. All points on this (hyper)surface--that will be a curve, for one degree of freedom--are equivalent. This means that, within the microcanonical ensemble, defined by this relation, all physical quantities are obtained  by taking averages over the ``angular variables'' of the constant energy (hyper)surface:
\begin{equation}
\label{vevsmicro}
\left\langle\mathcal{O}\right\rangle = \frac{\int\,dpdq\,\mathcal{O}\delta(H(p,q)-E)}{\int\,dpdq\,\delta(H(p,q)-E)}
\end{equation}
So for the simple harmonic oscillator, $H(p,q)=(p^2+q^2)/2$ and we are immediately incited to use ``polar'' coordinates, exchanging $(p,q)$ for $2E=p^2+q^2$ and $\theta=\tan^{-1}\,q/p$. In this way the denominator becomes equal to $2\pi$ and 
the numerator becomes equal to 
\begin{equation}
\label{numerO}
2\pi \left\langle\mathcal{O}(E)\right\rangle = \int_0^{2\pi}\,d\theta \mathcal{O}(E,\theta)\Leftrightarrow
\left\langle\mathcal{O}(E)\right\rangle = \int_0^{2\pi}\,\frac{d\theta}{2\pi}\,\mathcal{O}(E,\theta)\equiv 
\int_0^{2\pi}\,d\theta\,\rho(\theta)\,\mathcal{O}(E,\theta)
\end{equation}
In this case the group is $SO(2)\subset SL(2,\mathbb{R})$, a compact subgroup. We remark that $2\pi$ is the ``volume of the (hyper)surface'' in this case and this expression makes sense, because this number is finite; $\rho(\theta)$ is normalizable.

If we repeat the calculation for the case of the ``inverted oscillator'', where $H(p,q)=(p^2-q^2)/2$, though, we find that we are moving along a hyperbola, instead of on a circle, in phase space, so  when we perform the change of variables to the branch of the hyperobla,
$p=\sqrt{2E}\cosh\chi,q=\sqrt{2E}\sinh\chi$,  
\begin{equation}
\label{hyperbolic}
\left\langle\mathcal{O}(E)\right\rangle = \frac{\int_{-\infty}^\infty\,d\chi\,\mathcal{O}(E,\chi)}{\int_{-\infty}^{\infty}\,d\chi} =\int_{-\infty}^\infty\,d\chi \rho(\chi)\mathcal{O}(E,\chi)
\end{equation}
we find that we have a problem: instead of $2\pi$, we get infinity in the denominator; even though the energy is finite. The problem is that the constant energy manifold, of the group $SO(1,1)\subset SL(2,\mathbb{R})$, has infinite volume and it's not possible to apply a cutoff on it that can be removed--any quantity will be affected by its presence.  The density $\rho(\chi)$ isn't normalizable and can't be reconstructed from the moments.  It's this incompleteness that makes the microcanonical ensemble ill--defined in this case and makes any coupling to a bath, i.e. the canonical ensemble, problematic.  Stated differently, the change of variables from ``Cartesian'' to ``polar'' coordinates doesn't resolve the Hamiltonian constraint; it can't account for the states at infinity; and the reason it can't is that the particle undergoes accelerated motion, without bound.

In the literature this instability is conflated with the so--called Ostrogradsky instability~\cite{Ostrogradsky:1850fid}, that arises when the constant energy (hyper)surface is non--compact, because of multi--linear terms in the Hamiltonian. 

In both cases the problem is that the particle seems to undergo unbounded, accelerated motion. One way to resolve this is to identify the multilinear terms as describing the interaction of the particle with external fields. For the inverted oscillator, for instance, one can replace the $-q^2/2$ term by $(F^2/2)-Fq$ and recognize that the appropriate description is the interaction of the particle with an external field (a point that appears to have been overlooked in the comprehensive review~\cite{Barton:1984ey}.  Similarly, a term that's linear in the momentum can be absorbed in the redefinition of the momentum and its coefficient recognized as a vector potential (the symmetric part is a total derivative). 

However there is a remarkable difference between a linear term in momentum, from a linear term in position: the former describes, in fact, bounded motion perpendicular to the magnetic field and uniform motion, that may be eliminated by a Galilean transformation, along it; whereas the latter describes non--uniform, unbounded motion, along the electric field; that can't be eliminated by a Galilean transformation. The former describes the motion of the particle in a ``flux background''~\cite{Andriot:2012wx,Ntokos:2016vnm}; the latter describes the motion of a particle in ``geometrical'' background, that can be described through an ``Eisenhart lift''~\cite{Galajinsky:2016zer,Chanda:2016aph,Cariglia:2015bla}. These facts were, of course, not known in Ostrogradsky's time; but they are now. 

It is worth mentioning that, when working in phase space, which variable(s) are position(s) and which are momenta represents a choice of coordinates. Therefore the corresponding fields must transform accordingly under symplectic transformations; this has attracted the attention it deserves only recently under the heading of ``double field theory''~\cite{Hull:2009mi,Aldazabal:2013sca,Geissbuhler:2013uka,Bergshoeff:2016ncb}.

These redefinitions work when the equations of motion contain two time derivatives. Things become more complicated, when they contain higher derivatives. In that case terms that are multilinear in the dynamical variables appear in the integrals of motion in a way that seems to indicate that these don't define compact regions in phase space--illustrating the Ostrogradsky instability. But how to resolve it doesn't appear as obvious as in the case of systems with two derivatives, since the interpretation of the equations as describing the motion of individual particles is problematic, since these are, typically, in interaction. 

  So the question arises, whether it is possible to impose constraints that will lead to systems with compact constant energy surface, therefore leading to a situation where the energy is bounded by a universal constant. The activity in this area has focused on introducing constraints that eliminate the higher derivatives altogether and, in particular, in ways that lead to compact constant energy (hyper)surfaces~\cite{Govaerts:1994zh,Nakamura:1995qz, Morozov:2007rp,Chen:2012au,Langlois:2015cwa,Motohashi:2016ftl,Crisostomi:2017aim,Motohashi:2017eya}.
  In this way the higher derivatives are understood as coordinate artifacts; so, if their instabilities are absent, so are the benefits, accrued from their presence, namely improved behavior of the systems at short distances. 

Further applications to field theories, in particular, gravity~\cite{Galajinsky:2016zer,Langlois:2015cwa,Motohashi:2016ftl,Motohashi:2017eya,Bonanno:2013dja,Mohrbach:2000qc,Ketov:2011re}
has relied on intuition acquired from studying particle systems with higher derivatives, e.g.~\cite{Langlois:2015cwa,Motohashi:2016ftl,Crisostomi:2017aim,Motohashi:2017eya,Motohashi:2018pxg},
so it's worthwhile to study these from another point of view. 

There's been considerable effort, also, devoted to trying to define the Euclidian path integral~\cite{Hawking:1985gh,Hawking:2002ty,andrzejewski2010modified,andrzejewski2011euclidean,gosselin2002renormalization,Pavsic:2016ykq}, in terms of single particle states, with local dynamics, without, however, a conclusive result. 

The reason, in summary, is that, while it is, of course, possible to describe the dynamics in terms of a single particle, the dynamics won't be, necessarily, local, since the target space is two--dimensional and dimensional reduction to one dimension can't be implemented. That's the real issue and that's why these attempts haven't succeeded. 

Of course it shouldn't be surprising that equations of motion of even degree greater than two  in the worldvolume variables can describe many particles--or fields--in interaction:  By definition it's possible to express a differential equation of degree $2n$ as a system of $n$ differential equations, of degree two. The question is how can ``sensible'' interactions, described at the level of equations of second order,  be repackaged in terms of equations of higher order--and vice versa. Or, whether time--dependent solutions, consistent with the constraints, can be found.

The prototype is the so--called Pais--Uhlenbeck oscillator~\cite{Pais:1950kq}, that's described by a linear equation, of fourth order. The question, in view of the introductory remarks, is whether the phase space, defined by the constants of motion, can be compact or non--compact. If it is non--compact, then the system is open. if it is compact, then the system is consistently closed in the microcanonical ensemble. We shall find that there do exist parameter values for which the latter situation is relevant: the Pais--Uhlenbeck system can be consistently closed in the microcanonical ensemble, without any constraints, that have been discussed in the literature. 

However the consistently closed system, generically,  doesn't describe one particle, but two. In other words, the target space is, generically, two--dimensional.

We shall work in real time, since we  wish to investigate the microcanonical ensemble, as a prelude towards understanding what the correct way of defining the canonical ensemble might be; and we're interested in the space of solutions of the classical equations of motion.

The (classical) action of the Pais--Uhlenbeck oscillator is given by the following expression
\begin{equation}
\label{PU}
S_\mathrm{PU}[\phi]=\int\,dt\,\left\{\frac{1}{2}\dot{\phi}^2-\frac{\Omega^2}{2}\phi^2+\frac{\Gamma}{2}\ddot{\phi}^2\right\}
\end{equation}
This action is, for the moment, simply a way for defining an abstract variational problem--whether it is the action of one particle or many particles (or whether it can't be assigned to any physical system at all) remains to be understood. 

Whatever physical meaning may be attached to the function $\phi(t)$, what is unambiguous is that extremization of this functional of $\phi(t)$ implies solving the corresponding Euler--Lagrange equation 
\begin{equation}
\label{eomPU}
\Gamma\frac{d^4\phi}{dt^4}-\frac{d^2\phi}{dt^2}-\Omega^2\phi=0
\end{equation}
obtained the usual way:
\begin{equation}
\label{deltaS}
\delta S=\left.\left\{\delta\phi(\dot{\phi}-\Gamma\stackrel{\cdots}{\phi})+\delta\dot{\phi}\Gamma\ddot{\phi} \right\} \right|_{t_i}^{t_f}+\int_{t_i}^{t_f}\,dt\,\delta\phi\left\{\Gamma\frac{d^4\phi}{dt^4}-\frac{d^2\phi}{dt^2}-\Omega^2\phi\right\}
\end{equation}

To eliminate the boundary terms,  one must impose that 
\begin{equation}
\label{boundaryconditions}
\left.\left\{ \delta\phi(\dot{\phi}-\Gamma\stackrel{\cdots}{\phi})+\delta\dot{\phi}\Gamma\ddot{\phi}  \right\}\right|_{t_i}^{t_f}=0
\end{equation}
It is quite straightforward, when this boundary term vanishes,  to find two conserved quantities, that we shall call $E_1$ and $E_3$. They can be found by multiplying the equation of motion by $d\phi/dt$ and $d^3\phi/dt^3$ respectively:
\begin{equation}
\label{conserved}
\begin{array}{l}
\displaystyle
\dot{\phi}\left(\Gamma\frac{d^4\phi}{dt^4}-\frac{d^2\phi}{dt^2}-\Omega^2\phi\right)=0\Leftrightarrow
%\frac{d}{dt}\left(  \frac{1}{2}\left(\dot{\phi}^2+\Omega^2\phi^2\right) -\Gamma\left(\frac{d\phi}{dt}\frac{d^3\phi}{dt^3}-\frac{1}{2}
%\left(\frac{d^2\phi}{dt^2}\right)^2\right) \right)=0\Leftrightarrow \\
%\displaystyle
E_1 = \frac{1}{2}\left(\dot{\phi}^2+\Omega^2\phi^2\right) -\Gamma\left(\frac{d\phi}{dt}\frac{d^3\phi}{dt^3}-\frac{1}{2}
\left(\frac{d^2\phi}{dt^2}\right)^2\right)
\\
\displaystyle
\frac{d^3\phi}{dt^3}\left( \Gamma\frac{d^4\phi}{dt^4}-\frac{d^2\phi}{dt^2}-\Omega^2\phi\right)=0\Leftrightarrow
%\displaystyle
 E_3 =  
% \frac{1}{2}\Gamma\left(\frac{d^3\phi}{dt^3}\right)^2-\frac{1}{2}\ddot{\phi}^2-\Omega^2\phi\ddot{\phi}+\frac{\Omega^2}{2}\dot{\phi}^2=
% \frac{1}{2}\Gamma\left(\frac{d^3\phi}{dt^3}\right)^2-\frac{1}{2}\left(\ddot{\phi}^2+2\Omega^2\ddot{\phi}\phi+\Omega^4\phi^2-\Omega^4\phi^2\right)+\frac{\Omega^2}{2}\dot{\phi}^2=\\
% \displaystyle
% \hskip1truecm
 \frac{1}{2}\Gamma\left(\frac{d^3\phi}{dt^3}\right)^2-\frac{1}{2}\left(\ddot{\phi}+\Omega^2\phi\right)^2+\frac{\Omega^2}{2}\left(\dot{\phi}^2+\Omega^2\phi^2\right)
 \\
\end{array}
\end{equation}
With these expressions available, we can compute the microcanonical partition function,
\begin{equation}
\label{dos}
Z(E_1,E_3;\Omega,\Gamma)=\int\,du_0\,du_1\,du_2\,du_3\,\delta(E_1-E_1(u_0,u_1,u_2,u_3))\delta(E_3-E_3(u_0,u_1,u_2,u_3))
\end{equation}
where $u_I\equiv\phi^{(I)}(t)$, the successive derivatives of $u_0\equiv\phi(t)=\phi^{(0)}(t)$. If the two hypersurfaces intersect, in a surface of finite area, then this partition function is finite and the microcanonical averages have a chance of being well--defined. If they don't intersect, or they intersect along a non--compact surface,  the partition function either vanishes or diverges and the microcanonical averages don't exist. That there are two conserved quantities  has been overlooked in the literature to date. 

The partition function, thus,  depends on four variables: the two conserved charges and the parameters, $\Omega$ and $\Gamma$. 

If $\Gamma=0$, $E_1$ and $E_3$ are linearly dependent, since, in that case,  $\ddot{\phi}=-\Omega^2\phi$, therefore, 
\begin{equation}
\label{Gamma=0E1E3}
E_3(\Gamma=0)=\Omega^2 E_1(\Gamma=0)
\end{equation}
If we choose $\Omega^2 > 0$, $\Gamma < 0$ and $-4\Omega^2\Gamma < 1$, which means that $-1 < 4\Omega^2\Gamma < 0$, then the roots of the characteristic equation
\begin{equation}
\label{chareq}
\Gamma\lambda^4-\lambda^2-\Omega^2=0\Leftrightarrow\lambda=\pm\sqrt{\frac{1}{2\Gamma}\left(1\pm\sqrt{1+4\Omega^2\Gamma}\right)}
\end{equation}
 are pure imaginary, $\pm\mathrm{i}\omega_{1,2}$, with $\omega_{1,2}$ real, therefore the general solution is 
\begin{equation}
\label{solPU}
\phi(t)=K_1 e^{\mathrm{i}\omega_1 t} + K_2 e^{-\mathrm{i}\omega_1 t} + K_3 e^{\mathrm{i}\omega_2 t} + K_4 e^{-\mathrm{i}\omega_2 t}
\end{equation}
and describes a bounded function, for all finite values of the  constants--as long as $\omega_1\neq\omega_2$.  It doesn't describe runaway solutions, like the inverted oscillator. However,  what matters isn't so much the expression of the solution, but the properties of the conserved charges that it describes, that we shall study presently. 

While the individual constants, $K_I$ aren't constrained to be real, $\phi(t)$, of course, is. This means that the constants can be redefined so that $\phi(t)$ can be written as 
\begin{equation}
\label{solrealPU}
\phi(t)=A_1\cos\omega_1 t + B_1\sin\omega_1 t + A_2\cos\omega_2 t + B_2\sin\omega_2 t
\end{equation}
with the coefficients, $A_I$ and $B_I$ real. The ratios of these coefficients define the relative phase--there's only one, by time translation invariance, but it will be more convenient when computing the partition function to use this parametrization.

If $-4\Omega^2\Gamma=1$, then $\omega_1=\omega_2=\Omega\sqrt{2}$.  This is the ``degenerate'' case~\cite{Langlois:2015cwa,Motohashi:2016ftl,Crisostomi:2017aim,Motohashi:2017eya,Motohashi:2018pxg}. 
The equation, in this case, takes the form
\begin{equation}
\label{eqPUdeg}
\left(\frac{d^2}{dt^2}+2\Omega^2\right)^2\phi(t)=0
\end{equation}
whose general solution is given by the expressions
\begin{equation}
\label{solPUdeg}
\begin{array}{l}
\displaystyle
\phi(t)=A_1\cos\Omega\sqrt{2} t + B_1\sin\Omega\sqrt{2}t + t\left(A_2\cos\Omega\sqrt{2}t + B_2\sin\Omega\sqrt{2}t\right)\\
\end{array}
\end{equation}
We remark that this doesn't describe a bounded function, unless $A_2=0$ and $B_2=0$. However this doesn't, necessarily, mean that it doesn't describe a physical system, as we shall see, when we compute the expressions for $E_1$ and $E_3$. 

Another special case is when one, at least,  of the frequencies vanishes, i.e. $\Omega=0$. In that case the equation of motion takes the form
\begin{equation}
\label{Omega=0}
\Gamma\frac{d^4\phi}{dt^4}-\frac{d^2\phi}{dt^2}=0\Leftrightarrow\frac{d^4\phi}{dt^4}+\gamma^2\frac{d^2\phi}{dt^2}=0
\end{equation}
where we've set $\gamma^2\equiv-1/\Gamma>0$. 

We realize that, by setting $d^2\phi/dt^2\equiv\chi(t)$, we obtain a second order equation for $\chi(t)$, whose solution is 
\begin{equation}
\label{chisol}
\chi(t)=A\cos\gamma t + B\sin\gamma t
\end{equation}
therefore
\begin{equation}
\label{phisol}
\phi(t)=-\frac{A}{\gamma^2}\cos\gamma t -\frac{B}{\gamma^2}\sin\gamma t + Ct +D\equiv K\cos(\gamma t +\alpha)+Ct+D
\end{equation}
We can eliminate $D$ by time translation invariance, as will be confirmed from the expressions for the conserved quantities, $E_1$ and $E_3$; it doesn't seem that we can eliminate $C$, however. 

So what we shall do in the following  is compute the partition function, for the solutions obtained--and check, whether the point(s) in phase space, that describe one harmonic oscillator, alone, do belong to the admissible states or not (once it's established that these exist at all)--in other words, whether the integration over the complementary variables (a) factorizes and (b) is finite. If either condition isn't meant the ``unwanted'' mode doesn't decouple. 

To this end, we shall express the conserved quantities, $E_1$ and $E_3$, in terms of the integration constants and integrate over the latter. 

A straightforward  (though lengthy) calculation leads to the following expressions for $E_1$ and $E_3$ in terms of the constants, in the three cases studied:
\begin{itemize}
\item $-1 < 4\Omega^2\Gamma < 0$:
\begin{equation}
\label{E1andE3w1neqw2}
\begin{array}{l}
\displaystyle
E_1 = -\frac{\omega_1^2-\omega_2^2}{2(\omega_1^2+\omega_2^2)}\left( (A_1^2+B_1^2)\omega_1^2-(A_2^2+B_2^2)\omega_2^2\right)
\\
\displaystyle
E_3 = -\frac{\omega_1^2-\omega_2^2}{2(\omega_1^2+\omega_2^2)}\left( (A_1^2+B_1^2)\omega_1^4-(A_2^2+B_2^2)\omega_2^4\right)
\\
\end{array}
\end{equation}
These relations define {\em two} hyperbolas, that intersect in four points, since their asymptotes aren't identical, generically--as long as $E_1E_3\neq 0$ and $\omega_1\neq\omega_2$. 

These four points define the physical states of the system, i.e. the motion of the two oscillators. 

This becomes very clear, if we set  $x_1^2=A_1^2+B_1^2$ and $x_2^2=A_2^2+B_2^2$,  
in the $(A_1,B_1),(A_2,B_2)$ pairings. We easily solve for $x_1^2=H_1(E_1,E_3)$ and $x_2^2=H_2(E_1,E_3)$. Now $x_1$ and $x_2$ need not be non--negative. They just need to be real.

The microcanonical partition function is given by the expression
\begin{equation}
\label{microZw1neqw2}
\begin{array}{l}
\displaystyle
Z=\int\,dA_1dB_1dA_2dB_2\,\delta(E_1-E_1(A_1,B_1,A_2,B_2))\delta(E_3-E_3(A_1,B_1,A_2,B_2))=\\
\displaystyle
\int\,dA_1\,dB_1\,\delta(A_1^2+B_1^2-H_1(E_1,E_3))\int\,dA_2\,dB_2\,\delta(A_2^2+B_2^2-H_2(E_1,E_3))=\pi^2
\end{array}
\end{equation}
for the values of $E_1$ and $E_3$  that lead to intersecting hyperbolas, i.e. for which $H_1(E_1,E_3)\geq 0$ and $H_2(E_1,E_3)\geq 0$. The factorization doesn't mean that the oscillators decouple, since the constraints do depend on both charges $E_1$ and $E_3$. 

\item $-4\Omega^2\Gamma = 1\Leftrightarrow \omega_1=\omega_2=\omega$: (the ``degenerate'' case): In this case, $\Omega^2=\omega^2/2$ and 
$\Gamma=-1/(2\omega^2)$.  The corresponding expressions for $E_1$ and $E_3$ read 
\begin{equation}
\label{E1andE3w1=w2}
\begin{array}{l}
\displaystyle
E_1=-(A_2^2+B_2^2)-\omega(A_2B_1-A_1B_2)
\\
\displaystyle
E_3=  -2(A_2^2+B_2^2)\omega^2-\omega^3(A_2B_1-A_1B_2)
\\
\end{array}
\end{equation}
This case is of particular interest, since the function $\phi(t)$, given by~(\ref{solPUdeg}) isn't bounded--and the term that isn't bounded doesn't seem to represent a coordinate transformation that is an isometry of the two--derivative action of the bounded motion. 

However, appearances are deceiving!

We readily find that 
\begin{equation}
\label{rhoA2B2}
E_1\omega^2-E_3 =\omega^2(A_2^2+B_2^2)\Leftrightarrow A_2^2+B_2^2=E_1-\frac{E_3}{\omega^2}
\end{equation}
which implies that 
\begin{equation}
\label{A1B2A2B1}
A_1B_2-A_2B_1 = \frac{E_1+A_2^2+B_2^2}{\omega}=\frac{2E_1-\frac{E_3}{\omega^2}}{\omega}=\frac{2E_1}{\omega}-\frac{E_3}{\omega^3}
\end{equation}
and we recognize the definitions  of energy and of angular momentum of a particle moving on a plane--this is nothing more than motion in a uniform magnetic field, written in a very complicated way! This system describes a {\em single} particle--but in the presence of a flux, that can't be gauged away, if $2E_1\omega^2-E_3\neq 0$, so the particle explores a two--dimensional target space.   The two constraints leave two zeromodes, that describe the center of the particle's circular motion. 

Incidentally, we notice that, while $E_1$ and $E_3$ are independent constants of motion, they must satisfy the relation $E_1\omega^2\geq E_3$, for a solution to exist. The equality implies that $A_2=0=B_2$, and the second condition, then,  implies that $0=2E_1\omega^2-E_3$. Both can only be satisfied if $E_1=0=E_3$, which leaves $A_1$ and $B_1$ arbitrary, thereby describing a free, physical, oscillator, since the flux that binds  it vanishes.

Indeed, $2E_1\omega^2-E_3=0\Leftrightarrow \Omega^2 E_1-E_3=0$ is consistent with  the condition that the higher derivative term is an artifact (cf. eq.~(\ref{Gamma=0E1E3}); and $E_1\omega^2\geq E_3$ is a ``BPS bound'', with equality describing, precisely, the fact that the flux doesn't bind the particle. 

Therefore, iff the BPS bound is saturated,  the higher derivative is a coordinate artifact, since the equation is reducible to an equation with two derivatives. If the bound  isn't saturated, the particle doesn't decouple from the flux. 
\item $\Omega=0$: We readily find that 
\begin{equation}
\label{E1andE3W2=0}
\begin{array}{l}
\displaystyle
E_1=-\frac{A^2}{2}-\frac{B^2}{2}\\
\displaystyle
E_3=\frac{E_1}{\gamma^2}+\frac{C^2}{2}
\end{array}
\end{equation}
The partition function takes the form
\begin{equation}
\label{ZW2=0}
Z=\int\,dA\,dB\,dC\,\delta\left(2E_1+A^2+B^2\right)\delta\left(2(\gamma^2 E_3-E_1)-\gamma^2 C^2\right)
\end{equation}
which is finite and non-zero, as long as $E_1 \leq 0$ and $E_3\gamma^2-E_1\geq 0$. 
\end{itemize}
We remark that the expressions for each of the conserved quantities, $E_1$ and $E_3$ in the three cases, generically, are not of definite sign; but that's not relevant. What is relevant, for the microcanonical partition function, is that the intersection of the two surfaces be non--zero and finite. 

In conclusion, we have found that the correct interpretation of the Pais--Uhlenbeck oscillator is that it describes  two-- not one--particle states, unless  the ``BPS bound'' is saturated, when $1+4\Omega^2\Gamma=0$,  or when $\Omega=0$, in which case it does describe just one particle--a single harmonic oscillator.  

If $-1<4\Omega^2\Gamma<0$,  the Pais--Uhlenbeck oscillator, when it describes a bounded function,  describes two, interacting oscillators, that are never free. The interaction can be described through their phase coherence, which implies that they are linked through a flux. 

Another way of describing the system, however, is as a single particle, in a background, that carries flux. The reason is that the solution is a sum of two terms, that each describe a harmonic oscillator; but each term isn't an isometry for the two--derivative action that describes the other.

In both cases, the appropriate quantum description is obtained, as usual, by replacing the coefficients, $A_I$ and $B_I$ by operators and imposing canonical commutation relations. The two oscillators described this way can be packaged, à la Schwinger, in an angular momentum algebra. The question that remains to be clarified is, whether the vacua, annihilated by each annihilation operator, are equivalent, or not. 

 For the degenerate case, $4\Omega^2\Gamma=-1$, quantization follows the route of Landau and Peierls.  
 
 Specifically, we may solve 
 \begin{equation}
 \label{A2B2w1=w2}
 A_2^2+B_2^2=E_1-\frac{E_3}{\omega^2}
 \end{equation}
in terms of annihilation and creation operators: $b=A_2+\mathrm{i}B_2,  b^\dagger=A_2-\mathrm{i}B_2$ and $a=A_1+\mathrm{i}B_1, a^\dagger=A_1-\mathrm{i}B_1$; in which case the relation
\begin{equation}
\label{A1B2A2B1w1=w2}
A_1B_2-A_2B_1 = \frac{2E_1}{\omega}-\frac{E_3}{\omega^3}
\end{equation}
implies that $a$ and $b$ can't annihilate the same state, therefore the vacua aren't equivalent, if the BPS bound isn't saturated. This is a hallmark of quantum systems, in  non--trivial gravitational backgrounds, since the work of Hawking~\cite{Hawking:1974sw} (cf. also the work of Unruh~\cite{PhysRevLett.46.1351} and ~\cite{Brout:1995rd} for a comprehensive review). What wasn't appreciated then was that non--trivial gravitational backgrounds could be described by fluxes, e.g. along the lines of~\cite{Andriot:2012wx,Ntokos:2016vnm}. 

When $-1<4\Omega^2\Gamma < 0$, we may, also, solve   eqs.~(\ref{E1andE3w1neqw2}),  in terms of four operators, 
\begin{equation}
\label{oscillators}
\begin{array}{l}
\displaystyle
A_1^2+B_1^2=H_1(E_1,E_3)\\
\displaystyle
A_2^2+B_2^2=H_2(E_1,E_3)
\end{array}
\end{equation}
and we may define, as usual, creation and annihilation operators, $a=A_1+\mathrm{i}B_1, a^\dagger=A_1-\mathrm{i}B_1$, $b=A_2+\mathrm{i}B_2,  b^\dagger=A_2-\mathrm{i}B_2$. 
 in terms of these. These operators define the dynamics of the quantum system completely and consistently. 

To show that the vacua, $|0\rangle_{a,b}$, defined by $a|0\rangle_a=0$ and $b|0\rangle_b=0$, aren't equivalent, it suffices to note that eqs.~(\ref{E1andE3w1neqw2}) can be written in terms of the number operators, $N_a\equiv A_1^2+B_1^2$ and $N_b\equiv A_2^2+B_2^2$; as long as $E_1$ and $E_3$ don't both vanish, it's not possible to have $N_a=0$ and $N_b=0$. If $E_1$ and $E_3$ both vanish, on the other hand, it's not possible that $N_a$ or $N_b$ be non--zero. So the only state of the system is the vacuum state, for both oscillators.

 For $\Omega=0$, there is but one relation, $A^2+B^2=H(E_1,E_3,C)$, therefore one oscillator, in a flat background. Nevertheless, a ``BPS bound'', also, exists in this case, since $E_1\gamma^2\geq E_3$. When the bound is saturated, $C=0$. 
 
 $C$ can, indeed, be identified with the velocity of the ``partner'' to the oscillator described by $(A,B)$. So this case seems to describe an oscillator, interacting with a free particle--that moves with a fixed velocity--prescribed by the initial conditions. 
   
It is in the  way described here that transition amplitudes and probabilities can be defined and, thus, the consistent path integral in the presence of quantum fluctuations can be defined.

These results  open up new perspectives for describing entangled qubits and quantum oscillators in flux backgrounds, that have become the subject of real experiments~\cite{2005PhRvB..71f4503M,PhysRevApplied.8.014004,Chiorescu,PhysRevLett.96.067003}.

{\bf Acknowledgements:} It's a pleasure to thank G. W. Gibbons and J. Iliopoulos for many discussions on the Pais--Uhlenbeck oscillator and S. Mukohyama and K. Noui for discussions on the constraints of higher derivative theories.

\bibliographystyle{utphys}
\bibliography{Ostrogradsky1D}

\end{document}